\documentclass[twocolumn,showpacs,preprintnumbers,amsmath,amssymb,prb,floatfix]{revtex4}

\usepackage{graphicx}
\usepackage{dcolumn}
\usepackage{bm}
\usepackage{epsfig}

\begin{document}

\title{Nonlinear response of Bloch electrons in infinite dimensions}

\author{V.~Turkowski}
\email{turk@physics.georgetown.edu}
\homepage{http://www.physics.georgetown.edu/~turk}
\affiliation{Department of Physics, Georgetown University,
             Washington, D.C. 20057}%
\author{J.~K.~Freericks}
\email{freericks@physics.georgetown.edu}
\homepage{http://www.physics.georgetown.edu/~jkf}
\affiliation{Department of Physics, Georgetown University,
             Washington, D.C. 20057}

\date{\today}

\begin{abstract}
The exact nonlinear response of noninteracting (Bloch) electrons is examined
within a nonequilibrium formalism on the infinite-dimensional hypercubic
lattice.  We examine the effects of a spatially uniform, but time-varying 
electric field (ignoring magnetic-field effects).  The electronic Green's
functions, Wigner density of states, and time-varying current are
all determined and analyzed.  We study both constant and pulsed electric fields,
focusing on the transient response region.  These noninteracting Green's 
functions are an important input into nonequilibrium dynamical mean field theory
for the nonlinear response of strongly correlated electrons.
\end{abstract}



\maketitle

\section{\label{sec:level1} Introduction}

The linear-response theory of Kubo\cite{kubo_1957} and 
Greenwood\cite{greenwood_1958} is an attractive approach to understand how 
electrons (in the solid state) interact with external electromagnetic fields.
It can be used (in principle) to calculate general linear-response functions 
in systems
that have arbitrarily strong electron correlations. Surprisingly, the 
linear-response regime for many bulk materials (especially for parabolic band
semiconductors\cite{sarker_davies_khan_wilkins_1986,khan_davies_wilkins_1987,%
lipavsky_khan_abdolsalami_wilkins_1991,lipavsky_khan_kalvova_wilkins_1991,%
lipavsky_khan_wilkins_1991}) and devices, holds for a wide range of
electric field strengths.

But there are a multitude of interesting nonlinear effects in electric fields.
Most electronic devices have a nonlinear current-voltage relation
(transistors, Josephson junctions, etc.) and there is wide interest in nonlinear
effects in bulk materials as well (since it is the nonlinearity that often
determines the ultimate performance).

Devices are also becoming smaller and smaller. Semiconductor processing line 
features are well below 100~nm, and there is significant research effort on
nanoscale devices. In the latter case, a potential difference of one volt 
produces an electric field on the order of $10^7$~V/cm for nanometer
scaled devices.  These fields are large enough for nonlinear effects 
to be important, if not critical, to determine the proper behavior in an
external field.  There also has been significant research performed on high
energy density pulsed laser experiments, where fields as high as 
$10^{10}$~V/cm can easily be attained over a short time scale.  In that
case, one drives the material out of equilibrium by the pulse, and
studies how it relaxes back to an equilibrium distribution (as a means
to determine relaxation times, etc.).

There are few theoretical approaches to nonlinear effects in 
solid-state systems. The formalism was developed independently by 
Kadanoff and Baym\cite{kadanoff_baym_1962} and Keldysh\cite{keldysh_1964,%
keldysh_1965} in the early 1960s (Baym~\cite{baym_2000} and 
Keldysh~\cite{keldysh_2003} have each written short historical accounts of their
discoveries).  These approaches include the effects of external fields
to all orders and typically use perturbation theory to determine the
effects of many-body interactions\cite{rammer_smith_1986}.  
In the 1980s, Wilkins and collaborators%
\cite{jauho_wilkins_1982,jauho_wilkins_1984,%
sarker_davies_khan_wilkins_1986,khan_davies_wilkins_1987,%
davies_wilkins_1988,%
lipavsky_khan_abdolsalami_wilkins_1991,lipavsky_khan_kalvova_wilkins_1991,%
lipavsky_khan_wilkins_1991}
spent much effort in developing these ideas further, and in examining nonlinear
responses in finite dimensions. Here we extend that work to infinite-dimensional
lattices, where we find many of the results for the electronic Green's functions
can be determined analytically.  Our formalism allows for an analysis of
steady-state effects (like the Wannier-Stark ladders\cite{wannier_1962}) 
and of transient effects
(like the response to a pulsed field).  These noninteracting Green's functions are
a necessary input to a complete nonlinear response dynamical mean field theory
for strongly correlated electrons.  We will present results for that
work in a separate publication.

The organization of this contribution is as follows: in Section II, we present 
the formalism for the nonlinear response, in Section III, we present our
numerical results, and in Section IV, we present our conclusions.

\section{Green's functions for Bloch electrons in an external
electric field}

The Hamiltonian for tight-binding electrons hopping on a hypercubic lattice
(in the absence of any external fields) is
\begin{equation}
\mathcal{ H}=-\sum_{ij}t_{ij}c^\dagger_ic_j 
-\mu\sum_{i}c^\dagger_ic_i
\label{eq: hamiltonian}
\end{equation}
where $t_{ij}$ is the Hermitian hopping matrix (chosen to 
be\cite{metzner_vollhardt_1989}
$t_{ij}=t^*/2\sqrt{d}$ for nearest neighbors as $d\rightarrow\infty$), 
and $\mu$ is the chemical potential.
We shall consider the case when this system is 
coupled to an external electromagnetic field.
An electromagnetic field is described by a scalar potential $\phi({\bf r},t)$
and a vector potential ${\bf A}({\bf r},t)$ via 
\begin{equation}
{\bf E}({\bf r},t)=-\nabla \phi({\bf r},t)-\frac{1}{c}\frac{\partial {\bf A}
({\bf r},t)}{\partial t}
\label{eq: efield}
\end{equation}
for the electric field,
with $c$ the speed of light.  We will use the Landau gauge where $\phi=0$
to perform our calculations, so the electric field is described solely by the 
vector potential. This provides a significant simplification of the formalism
for spatially uniform (but possibly time-varying) electric fields.

Unlike many time-dependent Hamiltonians, the effect of the vector potential
is not easily described by adding a time-dependent piece to the Hamiltonian
in addition to the time-independent piece in Eq.~(\ref{eq: hamiltonian}).
Instead, one uses the so-called Peierls' substitution \cite{jauho_wilkins_1984}
for the hopping matrix:
\begin{equation}
t_{ij}\rightarrow t_{ij}\exp\left [ -\frac{ie}{\hbar c}
\int_{{\bf R}_i}^{{\bf R}_j}
{\bf A}({\bf r},t)\cdot d{\bf r}\right ],
\label{eq: peierls}
\end{equation}
where ${\bf R}_i$ is the spatial lattice vector associated with lattice site
$i$ (and similarly for site $j$) and $e$ is the electric charge. Note that the
Peierls' substitution is a simplified semiclassical treatment of the
electromagnetic field (our vector potential is a classical, not quantum
field) and we are ignoring dipole (and multipole) transitions between
bands because we consider just a single-band model.
In this case, the Hamiltonian of the noninteracting electrons
coupled to an electromagnetic field becomes
\begin{equation}
\mathcal{H}(t)=\sum_{ij}t_{ij}
\exp \left [ -\frac{ie}{\hbar c}
\int_{{\bf R}_i}^{{\bf R}_j} {\bf A}({\bf r},t)\cdot d{\bf r}\right ]
c^\dagger_ic_j -\mu\sum_{i}c^\dagger_ic_i.
\label{eq: hamiltonian_field}
\end{equation}
The corresponding electric field becomes
\begin{equation}
{\bf E}({\bf r},t)=-\frac{1}{c}\frac{\partial {\bf A}({\bf r},t)}{\partial t}.
\end{equation}
We will choose the vector potential in such a way that either the
field is zero before $t=0$ and is then turned on, or the field becomes
asymptotically small as $t\rightarrow-\infty$ and it is adiabatically
switched on; in this way, the early time Hamiltonian is always given
by Eq.~(\ref{eq: hamiltonian}), and that will be used to establish
the initial thermal equilibrium.
The magnetic field has a complicated structure in infinite dimensions, 
because it involves the curl of the
vector potential, which would need to be defined correctly for the 
infinite-dimensional limit. Because we are interested in electric fields with
weak spatial dependence, we shall assume the associated magnetic field is small
enough that we can neglect it, even though we will allow the electric field
to vary in time.  This is an approximation, because
our electromagnetic fields no longer satisfy Maxwell's equations, unless
the field is uniform in space and constant in time. This condition can be 
relaxed, perhaps by using a gradient expansion for the weak spatial dependence 
of the fields\cite{jauho_wilkins_1984}, but such an approach is cumbersome
in infinite dimensions. From now on,
we neglect the spatial dependence of the vector potential (i.e., we
are considering only spatially uniform but time-varying electric fields). 

It is convenient to introduce a momentum-space representation for the
Hamiltonian, which becomes
\begin{equation}
\mathcal{H}(t)= \sum_{\bf k}[\epsilon({\bf k}-\frac{e{\bf A}(t)}{\hbar c})-\mu]
c^\dagger_{\bf k}c_{\bf k},
\label{eq: hamiltonian_field_momentum}
\end{equation}
with $c_{\bf k}=\sum_jc_j\exp[i{\bf R}_j\cdot{\bf k}]$ and 
$c^\dagger_{\bf k}=\sum_jc^\dagger_j\exp[-i{\bf R}_j\cdot{\bf k}]$. Note that the 
Hamiltonian in Eq.~(\ref{eq: hamiltonian_field_momentum}) is a special 
time-dependent Hamiltonian, because it commutes with itself for all
times $[\mathcal{H}(t),\mathcal{H}(t^\prime)]=0$, which greatly simplifies
the analysis of the time-dependent Green's functions developed below.

The expression for the
time-ordered single-particle Green's function is defined to be
\begin{equation}
g^{T}({\bf k},t,t')=-\frac{i}{\hbar} 
\langle \mathcal{T}(c_{\bf k}(t)c_{\bf k}^{\dagger}(t'))\rangle;
\label{eq: top_gf}
\end{equation}
because of the special time dependence of the Hamiltonian, this Green's
function can be determined analytically.  In Eq.~(\ref{eq: top_gf}), the
operators are expressed in a Heisenberg picture, where the time dependence
is $\mathcal{O}(t)=\exp[it\mathcal{H}(t)]\mathcal{O}\exp[-it\mathcal{H}(t)]$
with $\mathcal{H}(t)$ determined from Eq.~(\ref{eq: hamiltonian_field}),
the time ordering symbol $\mathcal{T}$ orders earlier times to the right
(with a change of sign when two Fermionic operators are interchanged),
and the angle brackets indicate a thermal averaging $\langle \mathcal{O}\rangle
={\rm Tr} [\exp(-\beta \mathcal{H})\mathcal{O}]/{\rm Tr} 
[\exp(-\beta \mathcal{H})]$, with $\beta=1/T$ the inverse temperature
and the Hamiltonian being the field free (early-time) Hamiltonian
from Eq.~(\ref{eq: hamiltonian}).
We directly solve for the Green's function by finding the time dependence
of the momentum-dependent creation and annihilation operators, and
then directly solve for the Green's function by taking the relevant
expectation values and traces\cite{jauho_wilkins_1984,schmidt_monien_2002,%
schmidt_2002}. The starting point is to calculate the time
dependence of the operators:
\begin{eqnarray}
\frac{d}{dt}c^\dagger_{\bf k}(t)&=&\frac{i}{\hbar}
[\epsilon({\bf k}-\frac{e{\bf A}(t)}{\hbar c})-\mu]c^\dagger_{\bf k} (t)\\
\frac{d}{dt}c_{\bf k}(t)&=&-\frac{i}{\hbar}[\epsilon({\bf k}-\frac{e{\bf A}
(t)}{\hbar c})-\mu] c_{\bf k}(t)
\end{eqnarray}
which can be integrated to give
\begin{eqnarray}
c^\dagger_{\bf k}(t)&=&\exp\left [\frac{i}{\hbar}
\int_{-\infty}^t[\epsilon({\bf k}-\frac{e{\bf A}(\bar t)}{\hbar c})-\mu] d\bar t \right ]
c^\dagger_{\bf k} \label{eq: eom1}\\
c_{\bf k}(t)&=&\exp \left [-\frac{i}{\hbar}
\int_{-\infty}^t[\epsilon({\bf k}-\frac{e{\bf A}(\bar t)}{\hbar c})-\mu] d\bar t 
\right ]c_{\bf k}.
\label{eq: eom2}
\end{eqnarray}

It is now easy to find the expression for the time-ordered
Green's function by inserting the time dependence from Eqs.~(\ref{eq: eom1})
and (\ref{eq: eom2}) into the definition of the Green's function in
Eq.~(\ref{eq: top_gf}) to yield
\begin{eqnarray}
g^{T}({\bf k},t,t^\prime)&=&-\frac{i}{\hbar}\theta  (t-t^\prime)
\exp\left [-\frac{i}{\hbar}\int^t_{t^\prime}
[\epsilon({\bf k}-\frac{e{\bf A}(\bar t)}{\hbar c})-\mu ] d\bar t \right ]
\nonumber\\
&\times& [1-f(\epsilon({\bf k})-\mu)]\nonumber\\
&+&\frac{i}{\hbar}\theta(t^\prime-t)
\exp\left [-\frac{i}{\hbar}\int^t_{t^\prime}
[\epsilon({\bf k}-\frac{e{\bf A}(\bar t)}{\hbar c})-\mu] d\bar t\right ]\nonumber\\
&\times& f(\epsilon({\bf k})-\mu),
\label{eq: gk}
\end{eqnarray}
since the averages satisfy $\langle c^\dagger_{\bf k}c_{\bf k}\rangle=
f(\epsilon({\bf k})-\mu)$
and $\langle c_{\bf k}c^\dagger_{\bf k}\rangle=[1-f(\epsilon({\bf k})-\mu)]$
with $f(x)=1/[1+\exp(\beta x)]$ being the Fermi-Dirac distribution, and
$\epsilon({\bf k})$ the band structure.

In infinite-dimensional calculations, it is often important to also
determine local properties, like the local Green's function 
[$g_{loc}=\sum_{\bf k}g({\bf k})$], or the
local density of states (DOS).  The Green's function in Eq.~(\ref{eq: gk})
depends on both $\epsilon({\bf k})$ and $\epsilon({\bf k}-e{\bf A}/\hbar c)$.
Hence, the summation over momentum cannot be performed simply by introducing an
integral over the noninteracting DOS.  Instead, the method of Mueller-Hartmann 
must be used\cite{mueller-hartmann_1989a,mueller-hartmann_1989b,%
mueller-hartmann_1989c}, to perform the integrations over the Brillouin
zone and to extract the leading contributions as $d\rightarrow\infty$.
The algebra is straightforward, but lengthy.  The final result is
\begin{eqnarray}
g^T_{loc}(t,t^\prime)&=&-\frac{i}{\hbar}\int d\epsilon[\theta(t-t^\prime)
-f(\epsilon-\mu)]\nonumber\\
&\times&\rho(\epsilon)
\exp \left [ -i\frac{\epsilon}{\hbar}\frac{1}{d}\sum_\alpha
\int^t_{t^\prime}d\bar t \cos \frac{eaA_\alpha(\bar t)}{\hbar c}\right ]
\nonumber \\
&\times&\exp \frac{t^{*2}}{4\hbar^2}\Bigr\{\left [\frac{1}{d}\sum_\alpha
\int^t_{t^\prime}d\bar t\cos \frac{eaA_\alpha(\bar t)}{\hbar c}\right ]^2
\nonumber \\
&-& \frac{1}{d}\sum_\alpha
\int^t_{t^\prime}d\bar t \int^t_{t^\prime}d\bar t^\prime
\cos \frac{ea\{A_\alpha(\bar t)-A_\alpha(\bar t^\prime)\}}{\hbar c}\Bigr \}
\nonumber\\
&\times&e^{i\mu (t-t^\prime)/\hbar},
\label{eq: gloc_general}
\end{eqnarray}
where $\alpha$ denotes the component of the vector potential and
$\rho(\epsilon)=\exp[-\epsilon^2/t^{*2}]/\sqrt{\pi}t^*a^d$ is the
noninteracting DOS (and $a$ is the lattice spacing).  Note that
in the limit ${\bf A}\rightarrow 0$, this reduces to the well-known 
noninteracting Green's function on a hypercubic lattice.

While the results of Eq.~(\ref{eq: gloc_general}) are completely general,
they are quite cumbersome for calculations, and it is useful to consider
some simpler limits.  The easiest case to evaluate, which is what we
consider for the remainder of this paper, is to examine the case where the
vector potential lies along the 
$(1,1,1,...)$ diagonal [${\bf A}(t)=A(t)(1,1,1,...)$]. This choice
simplifies the calculations significantly. 
In this case, the momentum-dependent Green's function in Eq.~(\ref{eq: gk})
depends on just
two macroscopic objects---the band structure $\epsilon({\bf k})$ and
an additional energy function $\bar\epsilon({\bf k})=-t^*
\lim_{d\rightarrow\infty}\sum_\alpha \sin(k_\alpha a)/\sqrt{d}$:
\begin{eqnarray}
g^{T}(\epsilon, \bar \epsilon,t,t^\prime)&=&\exp\Bigr [
-\frac{i}{\hbar}\int^t_{t^\prime}\{\epsilon ({\bf k})\cos \frac{eaA(\bar t)}{\hbar c}\nonumber\\
&+&\bar\epsilon ({\bf k}) \sin\frac{eaA(\bar t)}{\hbar c}\}d\bar t\Bigr ]
e^{i\mu(t-t^\prime)/\hbar} \nonumber\\
&\times&\left ( -\frac{i}{\hbar}\right )
\left [ \theta(t-t^\prime)-f(\epsilon-\mu) \right ].
\label{eq: g_mom_alt}
\end{eqnarray}
Hence the local Green's function can be found by integrating over a joint
density of states\cite{schmidt_2002}
\begin{equation}
\rho_2(\epsilon,\bar\epsilon)=\sum_k \delta[\epsilon-\epsilon({\bf k})]
\delta[\bar\epsilon-\bar\epsilon({\bf k})],
\end{equation}
which yields
\begin{equation}
g^{T}_{loc}(t,t^\prime)=\int d\epsilon \int d\bar\epsilon \rho_2(\epsilon,
\bar\epsilon)
g^{T}(\epsilon,\bar \epsilon ,t,t^\prime).
\label{gklocal}
\end{equation}
Using the techniques
of Mueller-Hartman\cite{mueller-hartmann_1989a,%
mueller-hartmann_1989b,mueller-hartmann_1989c} again,  
gives the following expression for the joint density of states:
\begin{equation}
\rho_2(\epsilon,\bar\epsilon)=\frac{1}{\pi t^{*2}a^d}
\exp(-\frac{\epsilon^2}{t^{*2}}-\frac{\bar\epsilon^2}{t^{*2}}).
\label{rho2}
\end{equation}
Substituting the joint density of states of Eq.~(\ref{rho2})
into Eq.~(\ref{gklocal}) and integrating over $\bar\epsilon$ gives the
final expression for the local Green's function: 
\begin{eqnarray}
g^{T}_{loc}(t,t^\prime)&=&-\frac{i}{\hbar}\int d\epsilon[\theta(t-t^\prime)
-f(\epsilon-\mu)]\nonumber\\
&\times&\rho(\epsilon)
\exp[-i\frac{\epsilon}{\hbar}\int^t_{t^\prime}d\bar t \cos
\frac{eaA(\bar t)}{\hbar c}]\nonumber\\
&\times&\exp[-t^{*2}\left (\int^t_{t^\prime}d\bar t \sin
\frac{eaA(\bar t)}{\hbar c}\right )^2/4\hbar^2]\nonumber\\
&\times&e^{i\mu(t-t^\prime)/\hbar}.
\label{gc}
\end{eqnarray}
Of course, the result in Eq.~(\ref{gc}) agrees with that of Eq.~(\ref{eq: gloc_general}) when the vector potential lies along the diagonal.

\section{Numerical results}

We begin by studying the current density of the system in the presence of
the electric field.  The current operator is determined by
the commutator of the polarization operator (defined by
$\Pi=\sum_i \textbf{R}_ic^\dagger_ic_i$) with the
Hamiltonian of the system. The expression for the $\alpha$-component of the
current-density operator has the following form:
\begin{equation}
{\bf j}_\alpha=\frac{eat^*}{\hbar\sqrt{d}}\sum_{\bf k} 
\sin\left ( {\bf k}_\alpha a- \frac{ea{\bf A}_\alpha(t)}
{\hbar c}\right ) c^\dagger_{\bf k}c_{\bf k}.
\end{equation}
The expectation value of the $\alpha$th component of the 
current can be easily calculated from the time-ordered Green's function 
in Eq.~(\ref{eq: gk}) in the limit $t^\prime\rightarrow t^+$:
\begin{eqnarray}
\langle{\bf j}_\alpha \rangle &=&
-i\frac{eat^*}{\sqrt{d}}\sum_{\bf k} 
\sin\left ( k_\alpha a- \frac{eaA_\alpha(t)}
{\hbar c}\right ) g^T({\bf k},t,t^+),
\nonumber \\
&=&-\frac{eat^{*2}  }{4d\pi\hbar}
\sin\left ( \frac{eaA_\alpha(t)}{\hbar c}\right )\int d\epsilon
\frac{d f(\epsilon-\mu)}{d\epsilon}\rho(\epsilon),
\end{eqnarray}
where the summation over momentum is performed the same way as before.
The total magnitude of the current density
is just $\sqrt{d}$ times this result, since each 
component along the diagonal is the same.
In the limit of low temperature, we perform a Sommerfeld expansion, which gives
\begin{equation}
\langle j(t)\rangle=\frac{eat^{*2}\rho(\mu)}{4\sqrt{d}\pi\hbar}
\sin\left ( \frac{eaA(t)}{\hbar c}\right )
\end{equation}
[with $A(t)$ the value of the vector potential for each component].  
Note that in the case of a constant field, $A(t)=-Ect\theta(t)$ is a linear 
function of $t$,
and the current is sinusoidal, even though the field is time-independent.
This is the well-known Bloch oscillation,\cite{ashcroft_mermin_1976}
with a frequency $\omega_{Bloch} =eaE/\hbar$.  Since we have no scattering, 
the system is a perfect conductor, but the periodicity of the lattice 
restricts the wavevector to lie in the first Brillouin zone which causes
the oscillatory current.

One can investigate a current-current correlation function to determine
a noise spectrum, but because the current is periodic, the noise profile
would be just two delta functions for a constant field, and we won't learn
anything interesting from examining the noise.

It is interesting to note, that the current is nonzero for the
case $A(t)=const$, which corresponds to the case of zero electric field.
This is a consequence of the fact that the vector potential
results in a shifting of the Fermi surface. In the case of an interacting
system this current will be destroyed by interparticle scattering. In our
case, a free-energy analysis will show that the lowest-energy state is the
one without any current.  There are a number of analogies of the
response of this system to the response of a superconductor (such as
an ac response to a dc field, the presence of current-carrying states
that do not disappear over time, etc.).  All of these results are artifacts
of the lack of scattering in the system.

To find the resistivity of the system, we consider the case
of a uniform static electric field (along the diagonal) of magnitude 
$E\sqrt{d}$, which is turned on at $t=0$, so that
${\bf A}(t)=-{\bf E}ct\theta(t)$ 
[$A_\alpha(t)=-Ect\theta(t)$], and the potential along a path 
$b(1,1,1,...)/\sqrt{d}$ is equal to $V=-Eb\sqrt{d}$
(the length $b$ is the distance over which we have a potential drop).
The expression for the Ohm's law in the form $V=jRa^{d-1}$ 
(current density multiplied by the resistance-area product), gives the following
expression for resistance-area product:
\begin{equation}
Ra^{d-1}=\frac{V}{j}=\frac{4\pi \hbar Edb}{eat^{*2}\rho(\mu)}\Big /
\sin\left ( \frac{eaEt}{\hbar }\right ).
\end{equation}
The resistivity is defined to be $1/b$ times the resistance-area product,
in the linear-response limit of $E\rightarrow 0$. Therefore,
\begin{equation}
\rho_{lin.~resp.}=\frac{4\pi \hbar^2 d}{e^2a^2t^{*2}\rho(\mu)}\frac{1}{t}.
\label{lr}
\end{equation}
This result is proportional to $d$, as it should be because the conductivity
is proportional to $1/d$ in infinite dimensions.  The correct resistivity is 
zero for a 
noninteracting system. Here we see that the  linear-response resistivity 
in Eq.~(\ref{lr}) goes to zero in 
the limit of large time $t\rightarrow\infty$.

Let us estimate the linear response resistance of the ballistic metal from 
the expression in Eq.~(\ref{lr}), 
which can be finite  because the linear-response resistance has a factor of 
$b/t$ in it.
For the ballistic metal the length $b$ over which the electrons have moved in 
the time $t$ should be $b =v_Ft$, with $v_F$ a suitable average of the Fermi 
velocity.  This gives the resistance
\begin{equation}
R_{lin.~resp.}=\frac{4\pi\hbar^2 v_F d}{e^2 a^{d+1}t^{*2}\rho(\mu)}.
\end{equation}
This expression corresponds to the Sharvin 
resistance\cite{sharvin_1965_russia,sharvin_1965} for a single-band model in
infinite dimensions.  In three dimensions, the Sharvin resistance is $h/2e^2$ 
divided by the number of channels, which
is a Fermi surface factor multiplied by $4\pi/k_F^2 Area$.  To compare
with our formula, we must first note that we map the hopping integral onto the
effective mass (for low electron filling) via
\begin{equation}
t^*=\frac{\hbar^2 \sqrt{d}}{m a^2}
\end{equation}
and that $a^dt^*\rho(\mu)=C$ is a constant of order one (proportional to
$(k_Fa)^{d-2}$ for low filling).  Therefore,
\begin{equation}
R_{lin.~resp.}=\frac{4\pi m v_F a \sqrt{d}}{e^2 C}
\propto\frac{h}{2e^2}\frac{4 \sqrt{d}}{(k_Fa)^{d-3}},
\end{equation}
which has a Sharvin-like form (but appears to have the wrong dependence
on $k_Fa$ for $d=3$; this most likely is an artifact of the problems
with assuming a spherical Fermi surface in large dimensions, which
is valid only for vanishing electron densities).

We can also investigate the heat current carried when there is an
electrical field present (but no temperature gradient), and we find
that its average value vanishes at half filling, as expected, because the
thermopower vanishes at half filling, and we have no thermal gradients
to directly drive a thermal current (in the general case, the energy part of
the current vanishes, and the chemical potential piece will give a
contribution of $-\mu {\bf j}$ to the heat current). So heat transport is
trivial unless one introduces a thermal gradient to the temperature, which
we do not do here.

Next we examine the spectral function and the density of states
in the presence of a field.
The time-dependent spectral function can be calculated from the retarded
Green's function $g^{R}(t,t^\prime)=-(i/\hbar )\theta(t-t^\prime)
\langle\{c(t),c^{\dagger} (t^\prime)\}\rangle$
(with the operators expressed in a Heisenberg picture)
using the Wigner coordinates\cite{wigner_1932} by introducing the average time
$t_{ave}=(t+t^\prime)/2$ and the relative time $t_{rel}=t-t^\prime$ variables.
In this case, the spectral function as a function of the average time 
(and Fourier transformed over the relative time) is equal to
\begin{equation}
A(t_{ave},{\bf k},\omega)=-\frac{1}{\pi}{\rm Im}
\int_{0}^{\infty} d t_{rel} e^{i\omega t_{rel}} g^R({\bf k},t_{ave},t_{rel}),
\label{Atk}
\end{equation}
and the DOS is equal to
\begin{equation}
A(t_{ave},\omega)=-\frac{1}{\pi}{\rm Im}
\int_{0}^{\infty} d t_{rel} e^{i\omega t_{rel}} g^R_{loc}(t_{ave},t_{rel}).
\label{At}
\end{equation}

In general, the retarded Green's function can be found from the 
same technique used to calculate the time-ordered Green's function:
first one introduces the time dependence of the Heisenberg operators,
then one evaluates the operator averages.  Since the anticommutator of two
local creation and annihilation operators (or two operators in the
momentum basis) is equal to one, we get
\begin{eqnarray}
g^{R}({\bf k},t,t^\prime)&=&-\frac{i}{\hbar}\theta(t-t^\prime)
e^{i\mu(t-t^\prime)/\hbar}
\nonumber\\
&\times& \exp[-i\frac{\epsilon({\bf k})}{\hbar}\int^t_{t^\prime}d\bar t \cos
\frac{eaA(\bar t)}{\hbar c}]\nonumber\\
&\times& \exp[-i\frac{\bar\epsilon({\bf k})}{\hbar}\int^t_{t^\prime}d\bar t \sin
\frac{eaA(\bar t)}{\hbar c}]
\label{gk_ret}
\end{eqnarray}
for the momentum-dependent Green's function and 
\begin{eqnarray}
g^{R}_{loc}(t,t^\prime)&=&-\frac{i}{\hbar}\theta(t-t^\prime)
\int d\epsilon \rho(\epsilon) e^{i\mu(t-t^\prime)/\hbar}
\nonumber\\
&\times& \exp[-i\frac{\epsilon}{\hbar}\int^t_{t^\prime}d\bar t \cos
\frac{eaA(\bar t)}{\hbar c}]\nonumber\\
&\times&\exp[-t^{*2}\left (\int^t_{t^\prime}d\bar t \sin
\frac{eaA(\bar t)}{\hbar c}\right )^2/4\hbar^2],
\label{gc_ret}
\end{eqnarray}
for the local Green's function (using the $t$ and $t^\prime$ coordinates). 
Note that these Green's functions have no temperature dependence, hence the
spectral function and the
DOS are independent of temperature.  This is characteristic of a noninteracting 
system.

The spectral function, in the absence of a field, is a delta function
[$A({\bf k},\omega)=\delta(\omega-\epsilon({\bf k})+\mu)$]. When a field
is turned on, the time dependence is no longer a pure exponential, so
the spectral function deviates from the delta function, becoming a peaked
function of nonvanishing width.  In the limit where $t_{ave}\rightarrow\infty$,
the steady state is approached and the spectral function becomes a set
of evenly spaced delta functions, since the Green's function becomes a periodic
function in $t_{rel}$.

The analysis for the local DOS is more complicated.
Since the $\epsilon$ dependence in Eq.~(\ref{gc_ret}) is so simple, the
integral can be performed directly, with the result
\begin{eqnarray}
g^R_{loc}(t_{ave},t_{rel})&=&-\frac{i}{\hbar}\theta(t_{rel})e^{i\mu t_{rel}/\hbar}\nonumber\\
&\times&\exp\left [ -\frac{t^{*2}} {4\hbar^2} |I(t_{ave},t_{rel})|^2\right ],
\end{eqnarray}
where 
\begin{equation}
I(t_{ave},t_{rel})=\int_{t_{ave}-t_{rel}/2}^{t_{ave}+t_{rel}/2}
d\bar t \exp\left [ i\frac{eaA(\bar t)}{\hbar c}\right ].
\label{eq: i_def}
\end{equation}

In order to evaluate some numerical results, we first consider the case
of a constant electric field turned on at $t=0$.  In this case, 
we get
\begin{eqnarray}
I(t_{ave},t_{rel})&=&\theta(-t_{ave}-t_{rel}/2)\theta(-t_{ave}+t_{rel}/2)
t_{rel}\nonumber\\
&+&\theta(-t_{ave}-t_{rel}/2)\theta(t_{ave}-t_{rel}/2)\nonumber\\
&\times&[t_{ave}+t_{rel}/2
+(1-e^{i\frac{eaE}{\hbar}(t_{ave}-t_{rel}/2)})\frac{\hbar}{ieaE}]\nonumber\\
&+&\theta(t_{ave}+t_{rel}/2)\theta(-t_{ave}+t_{rel}/2)\nonumber\\
&\times&[(e^{i\frac{eaE}{\hbar}(t_{ave}+t_{rel}/2)}-1)\frac{\hbar}{ieaE}
-t_{ave}+t_{rel}/2]
\nonumber\\
&+&\theta(t_{ave}+t_{rel}/2)\theta(t_{ave}-t_{rel}/2)\label{eq: i_def2}\\
&\times&\frac{\hbar}{ieaE}
(e^{i\frac{eaE}{\hbar}(t_{ave}+t_{rel}/2)}-e^{i\frac{eaE}{\hbar}
(t_{ave}-t_{rel}/2)})\nonumber.
\end{eqnarray}
This result has some interesting properties.  If $E\rightarrow 0$, then
$I=t_{rel}$ for all $t_{ave}$, and $g^R_{loc}$ is a Gaussian in $t_{rel}$, which
Fourier transforms to a Gaussian in frequency, i.e., it becomes the 
noninteracting DOS. There is an interesting scaling behavior.
If we define $\bar t_{ave}=t_{ave}eaE/\hbar$, $\bar t_{rel}=
t_{rel}eaE/\hbar$, and $\bar\omega=\omega \hbar/eaE$, then
\begin{equation}
I(t_{ave},t_{rel})=\frac{\hbar}{eaE}\bar I(\bar t_{ave},\bar t_{rel}),
\end{equation}
with $\bar I$ a function independent of $E$.  Hence
\begin{eqnarray}
g^R_{loc}(\bar t_{ave},\bar t_{rel})&=&-\frac{i}{\hbar}\theta(\bar t_{rel})
e^{i\mu \bar t_{rel}/eaE}\nonumber\\
&\times&\exp\left [
-\frac{t^{*2}}{4e^2a^2E^2}|\bar I(\bar t_{ave},\bar t_{rel})|^2\right ],
\label{eq: g_loc_scale}
\end{eqnarray}
and the DOS becomes
\begin{equation}
A(\bar t_{ave},\bar \omega)=-\frac{1}{\pi }{\rm Im}\int_0^\infty
d\bar t_{rel}e^{i\bar\omega\bar t_{rel}}g^R_{loc}(\bar t_{ave},\bar t_{rel}),
\end{equation}
with the normalization chosen so $\int d\bar\omega A(\bar\omega)=1$
(for easier comparison of curves for different $E$).
Hence we expect the DOS to have the same shape as a function of $\bar\omega$
(with a possible shift due to the chemical potential factor),
but the amplitude of the oscillations grows as $E$ increases [because of the
minus sign in the exponent in Eq.~(\ref{eq: g_loc_scale})]. But that turns
out only to be true near $\omega=0$.  At other frequencies, the evolution
with $E$ is not always monotonic, because the DOS conserves total spectral
weight, so there cannot be a monotonic evolution of the peaks at all
frequencies.

Note that the DOS satisfies two properties in equilibrium.  The first is
that the integral over frequency equals 1.  The second is that the DOS is
always positive.  The proof for the integral yielding 1 holds even in the
nonequilibrium case, because the anticommutator of two Fermionic creation
and annihilation operators at the same time is still one.  The positivity
does not hold, because the standard derivation, using the spectral
representation, requires the Hamiltonian to be independent of time in order
to be able to be used, and thereby prove the positivity. Indeed, the
DOS in the presence of a field has regions where it is negative.
 
\begin{figure}[h] 
\centering{
\includegraphics[width=8cm]{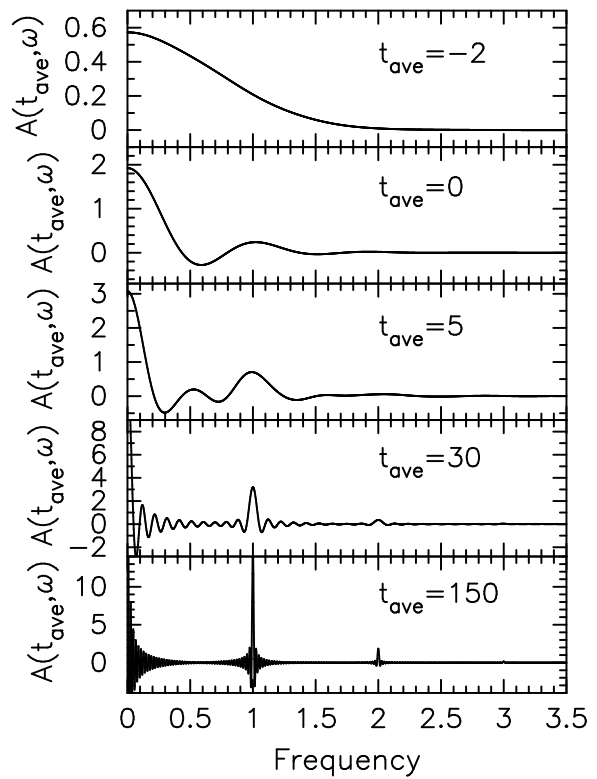}}
\caption{
Density of states $A(t_{ave},\omega)$ for the noninteracting electrons with
$eaE/\hbar=1$.  Note how the DOS is essentially a Gaussian for $t_{ave}<-2$,
but then develops large oscillations as $t_{ave}$ increases.  The DOS
approaches a steady state for large time given by a set of delta functions,
equally spaced by the Bloch oscillation frequency.
The DOS is no longer positive once the field is
turned on, but the integral does always equal 1.
}  
\label{fig: dos}
\end{figure}

It is interesting to consider the limit of large
$t_{ave}$, i.e., $t_{ave}\rightarrow\infty$, then we get the steady-state 
solution.  We take only the last term of $I(t_{ave},t_{rel})$ in 
Eq.~(\ref{eq: i_def2})
because $t_{ave}$ is always larger than $t_{rel}$ in this limit.  The Green's
function becomes
\begin{eqnarray}
g^R_{loc}(t_{ave}&\rightarrow&\infty,t_{rel})=
-\frac{i}{\hbar}\theta(t_{rel})\nonumber\\
&\times&
\exp\left [ \frac{t^{*2}}{2e^2a^2E^2}\left (
\cos(\frac{eaE}{\hbar}t_{rel})-1\right )\right ].
\end{eqnarray}
The Fourier transform of this is a set of delta functions, with different
amplitudes, that are equally spaced in frequency, with a spacing
$eaE/\hbar$ (since the Green's function is periodic in $t_{rel}$).
This is the famous Wannier-Stark ladder\cite{wannier_1962}, expected for
systems placed in an external
electric field.  In the results plotted in Fig.~\ref{fig: dos}, the fact that 
the peaks at multiples
of this frequency get larger, and grow in height as $t_{ave}$ grows,
indicates our results are showing the correct build-up to the steady
state, but they will never get there until $t_{ave}\rightarrow\infty$.
It is no coincidence that this frequency is the same as the Bloch oscillation
frequency.  This discussion was first described in detail from the
Green's function approach by Davies and Wilkins\cite{davies_wilkins_1988}.
Note that the DOS is nonnegative in the steady state.

We can calculate the weight of the delta functions by performing the Fourier
series integral.  The frequencies are $N eaE/\hbar$, and the Fourier coefficient
is
\begin{eqnarray}
w_N&=&\frac{2}{eaE}\int_0^{\frac{2\pi\hbar}{eaE}}dt_{rel}\cos(\frac{NeaE}
{\hbar}t_{rel})\nonumber\\
&\times&
\exp\left [ \frac{t^{*2}}{2e^2a^2E^2}\left (
\cos (\frac{eaE}{\hbar}t_{rel})-1\right )\right ]
\nonumber\\
&=&\frac{2\hbar}{e^2a^2E^2}\int_0^{2\pi}du\cos(Nu)
\exp(\frac{t^{*2}}{2e^2a^2E^2}
[\cos u-1]).
\end{eqnarray}

For our numerical results, we examine how the system approaches the steady
state as the field is turned on.  We work at half filling ($\mu=0$), where
the DOS is symmetric; hence, we plot only the results for positive
frequencies.  The field needs to be large enough for
our calculations to be able to see the nonlinear effects of the field
on the DOS.  For us, the numerical results can easily see effects on the DOS
when $eaE/\hbar>0.1$.  In Fig.~\ref{fig: dos}, we plot results for 
$eaE/\hbar=1$.  While it is
true that the Green's functions for $t$ and $t^\prime$ both less than zero
are equal to their equilibrium (field-free) limit, the Wigner DOS feels the 
effect of the fields for all finite $t_{ave}$, because the integral over 
$t_{rel}$ always includes some Green's functions with either $t$ or $t^\prime$
larger than zero.  We can see that significant ``precursor'' effects occur only
for $t_{ave}>-2$ here, and the DOS develops significant oscillations before
one can see the delta functions start to build up at the integer frequencies.

\begin{figure}[h] 
\centering{
\includegraphics[width=8cm]{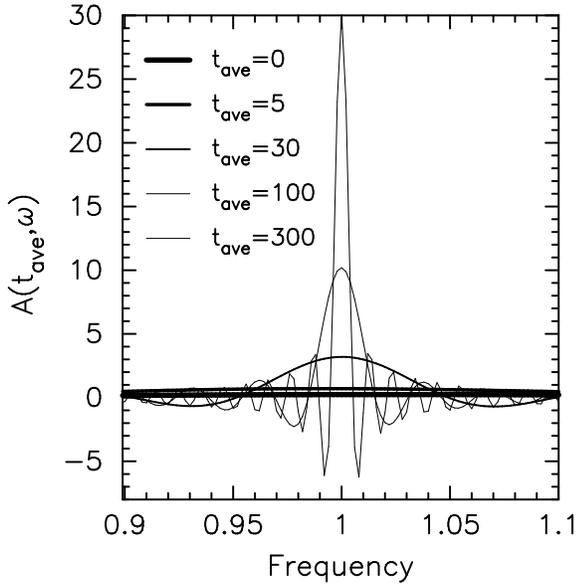}}
\caption{
Close up of the density of states $A(t_{ave},\omega)$ near $\omega=1$
for the noninteracting electrons with
$eaE/\hbar=1$.  Note how the DOS 
approaches a steady state for large time by developing a sharp peak,
but that there are significant oscillations near the sharp peak that
decay slowly in time.
}
\label{fig: dos2}
\end{figure}

We plot a close up of the region  around $\omega=1$ in Fig.~\ref{fig: dos2}.
Note how a sharp peak develops as the average time increases, but there are
significant oscillations near $\omega=1$ whose amplitude decreases slowly
as $t_{zve}$ increases.

\begin{figure}[h] 
\centering{
\includegraphics[width=8cm]{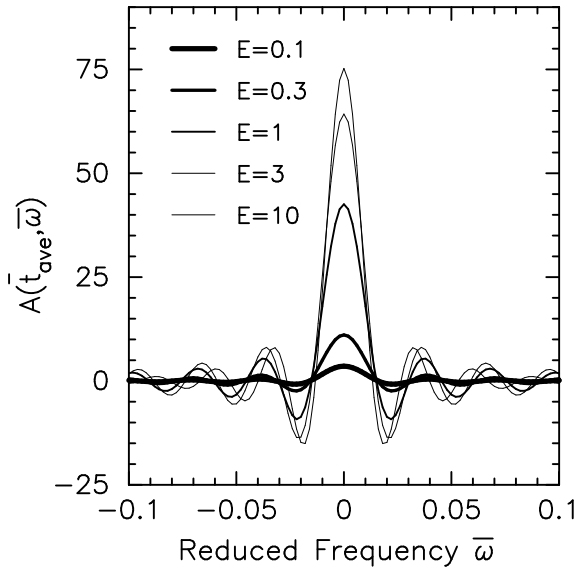}}
\caption{
Close up of the density of states $A(\bar t_{ave},\bar \omega)$ near $\bar
\omega=0$ for the noninteracting electrons with
$eaE/\hbar=0.1$, 0.3, 1.0, 3.0, and 10.0.  Note how the peak in the DOS
evolves as a function of the electric field.
}
\label{fig: dos3}
\end{figure}

In Fig.~\ref{fig: dos3}, we plot the DOS in the $\bar\omega$ variable
near $\bar\omega=0$ for $\bar t_{ave}=100$ and for five values of $eaE/\hbar$
(0.1, 0.3, 1.0, 3.0, and 10.0).  This shows how the oscillations grow as $E$
increases. For other integer values of $\omega$, the evolution is
not monotonic in the field strength $E$ (for example, at $\omega=1$ the peak 
values increase with $E$ for $0.1<eaE/\hbar <0.7$ and then decrease for
$0.7<eaE/\hbar <10$).

In addition to the spectral function and the DOS, it is interesting to examine 
the distribution function.  In equilibrium, the distribution function is
a Fermi-Dirac distribution function, but the distribution function can
change for nonequilibrium cases.  In order to discuss distribution functions, we
need to define two more Green's functions---the so-called lesser and greater
Green's functions.  They are defined as $g^{>}(t,t^\prime)=-(i/\hbar )\langle 
c(t) c^{\dagger}(t^\prime)\rangle$ and $g^{<}(t,t^\prime)=(i/\hbar )\langle 
c^\dagger(t^\prime) c(t)\rangle$
(with the operators expressed in a Heisenberg picture).  These Green's functions
can also be  determined exactly for Bloch electrons, and their expressions
are the same as those for the retarded Green's function in Eqs.~(\ref{gk_ret}) 
and (\ref{gc_ret}), except the $\theta(t-t^\prime)$ factor is replaced by
$-f(\epsilon({\bf k})-\mu)$ for $g^<$ and by $[1-f(\epsilon({\bf k})-\mu)]$ 
for $g^>$.  There are three cases for the distribution
function that we can consider (the Wigner distribution, the quasiparticle 
distribution, and the local quasiparticle distribution).  The most often used 
distribution function is the Wigner distribution function, defined to be
\begin{equation}
f_{Wigner}(t_{ave},{\bf k})=-i\hbar g^<({\bf k},t=t_{ave},t^\prime=t_{ave}).
\label{eq: wigner_def}
\end{equation}
The Wigner distribution function is always equal to the field-free Fermi-Dirac
result $f_{Wigner}(t_{ave},{\bf k})=f(\epsilon({\bf k})-\mu)$ for Bloch
electrons.  The quasiparticle distribution function is defined in analogy
with the equilibrium result [$g^<({\bf k},\omega)= 2\pi i f(\omega)
A({\bf k},\omega)$] via 
\begin{equation}
f_{quasi}(t_{ave},{\bf k})=\frac{1}{2\pi}\frac{{\rm Im}g^<(t_{ave},{\bf k},
\omega)} {A(t_{ave},{\bf k},\omega)}
\label{eq: quasi_def}
\end{equation}
(note that the name quasiparticle distribution does not necessarily imply
that there must be an underlying Fermi-liquid in the system).
Since the only difference between the retarded Green's function and
the lesser Green's function is the replacement of the theta function by
the Fermi-Dirac distribution (which does not depend on the time variables),
the ratio of the two terms in Eq.~(\ref{eq: quasi_def}) has an explicit factor 
of $f(\epsilon({\bf k})-\mu)$.  The Fourier transform of the numerator is over
all $t_{rel}$, while the denominator is only over all positive $t_{rel}$.
The integral $I(t_{ave},t_{rel})$ is an odd function of $t_{rel}$ [see 
Eq.~(\ref{eq: i_def})], which implies the numerator in Eq.~(\ref{eq: quasi_def})
is $2\pi f(\epsilon({\bf k})-\mu)A(t_{ave},{\bf k},\omega)$, and we find
the quasiparticle distribution function is equal to the Fermi-Dirac distribution
once again.  The final distribution function to be defined is the local
quasiparticle distribution function.  This is
\begin{equation}
f_{quasi}^{loc}(t_{ave})=\frac{1}{2\pi}\frac{{\rm Im}g^<_{loc}(t_{ave},
\omega)} {A(t_{ave},\omega)}.
\label{eq: loc_quasi_def}
\end{equation}
This distribution function is nontrivial in a field, because the
DOS and the lesser Green's function both have oscillations, but the zeros
occur at different locations on the frequency axis, so the ratio in 
Eq.~(\ref{eq: loc_quasi_def}) can have significant oscillations.

\begin{figure}[h] 
\centering{
\includegraphics[width=8cm]{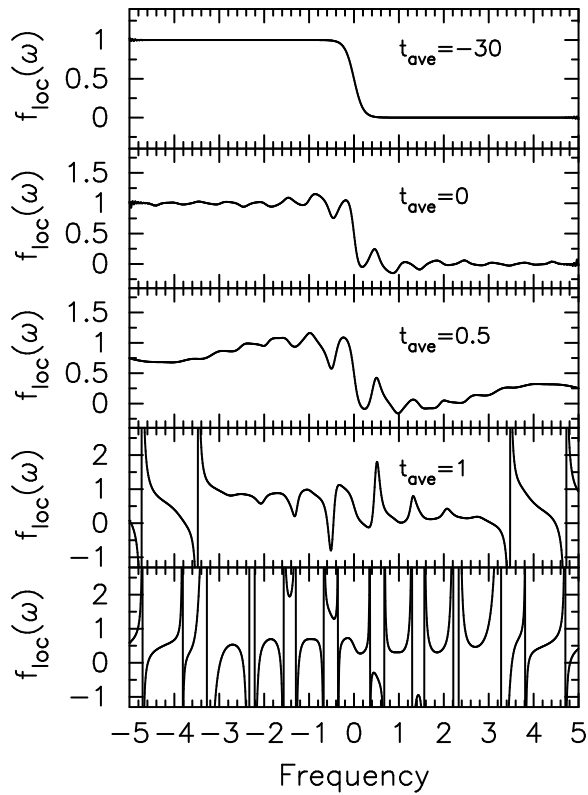}}
\caption{
Local quasiparticle
distribution function $f_{loc}(t_{ave},\omega)$ for noninteracting electrons 
with $eaE/\hbar=1$ and $T=0.1$.  Note how the local quasiparticle
distribution function varies
significantly from the equilibrium values as $t_{ave}$ increases
(the lowest panel is for $t_{ave}=2$).  This is 
because the $g^<$ Green's function has high
frequency oscillations, which are not as
strong in the DOS.  The oscillations continue as $t_{ave}$ increases, but they
become difficult to plot. 
}
\label{fig: dist}
\end{figure}

The calculation of the local quasiparticle distribution function is
difficult because the
presence of an $f(\epsilon-\mu)$ factor precludes us from performing the
integral over $\epsilon$ analytically; hence the numerical computations are
more involved. We need to evaluate the integral
\begin{eqnarray}
g^<(t_{ave},t_{rel})&=&\frac{i}{\hbar}\int d\epsilon \rho(\epsilon)
f(\epsilon-\mu)\nonumber\\
&\times&
\exp\left [-i\frac{\epsilon}{\hbar}x(t_{ave},t_{rel})-\frac{t^{*2}}{4\hbar^2}
y^2(t_{ave},t_{rel})\right ]
\end{eqnarray}
numerically.
If $eaE/\hbar=0$, then this is just the Fourier transform of $2\pi if(\omega)
\rho(\omega)$, which gives the correct lesser function.  If $eaE/\hbar\ne 0$, 
then the Green's function has to be calculated numerically.  Because the
real part of the lesser Green's function is nonzero for a longer range
in time than the imaginary part, the function $g^<$ will have more
oscillations than the $g^R$ function.  The results for a local quasiparticle
distribution function are plotted in Fig.~\ref{fig: dist}.
As it follows from this figure, the local quasiparticle
distribution function varies
significantly from the equilibrium values as $t_{ave}$ increases.  This is 
because the $g^<$ Green's function has high
frequency oscillations, which are not as
strong in the DOS.  The oscillations continue as $t_{ave}$ increases, but they
become difficult to plot. Of course the momentum-dependent quasiparticle
distribution function is equal to the Fermi-Dirac distribution function
for this problem.

Finally, we study the time dependence of the DOS
for the case of a sharp pulse during the period of time
$0<t<t_{E}$. The second derivative of the vector potential is proportional
to the strength of the magnetic field (which we are neglecting), so we
want to keep the second derivative small for the calculations to make
sense.  We choose the electric field to have the 
following time dependence: $E(t)=E\theta (t_{E}-t)\theta (t)$,
which corresponds to a vector potential
\begin{equation}
A(t)=-cEt\theta (t_{E}-t)\theta (t)-cEt_{E}\theta (t-t_{E}).
\end{equation}
Note that these results are ``singular'' for the noninteracting case, 
because the final vector potential is a constant that can correspond to
a current carrying state if the Fermi surface is shifted from the
zone center.  Because there is no scattering, such a current lives forever
(but would decay in the presence of any scattering).
Numerical calculations show that the DOS
deviates visibly from its equilibrium value during the times $|t|<t_{relax}$
when the amplitude of the field is larger or on the order
of $t^{*}$; the relaxation time $t_{relax}$ is on the order
of the pulse time $t_{E}$.

\begin{figure}[h!]
\centering
\includegraphics[width=3.0in,angle=0]{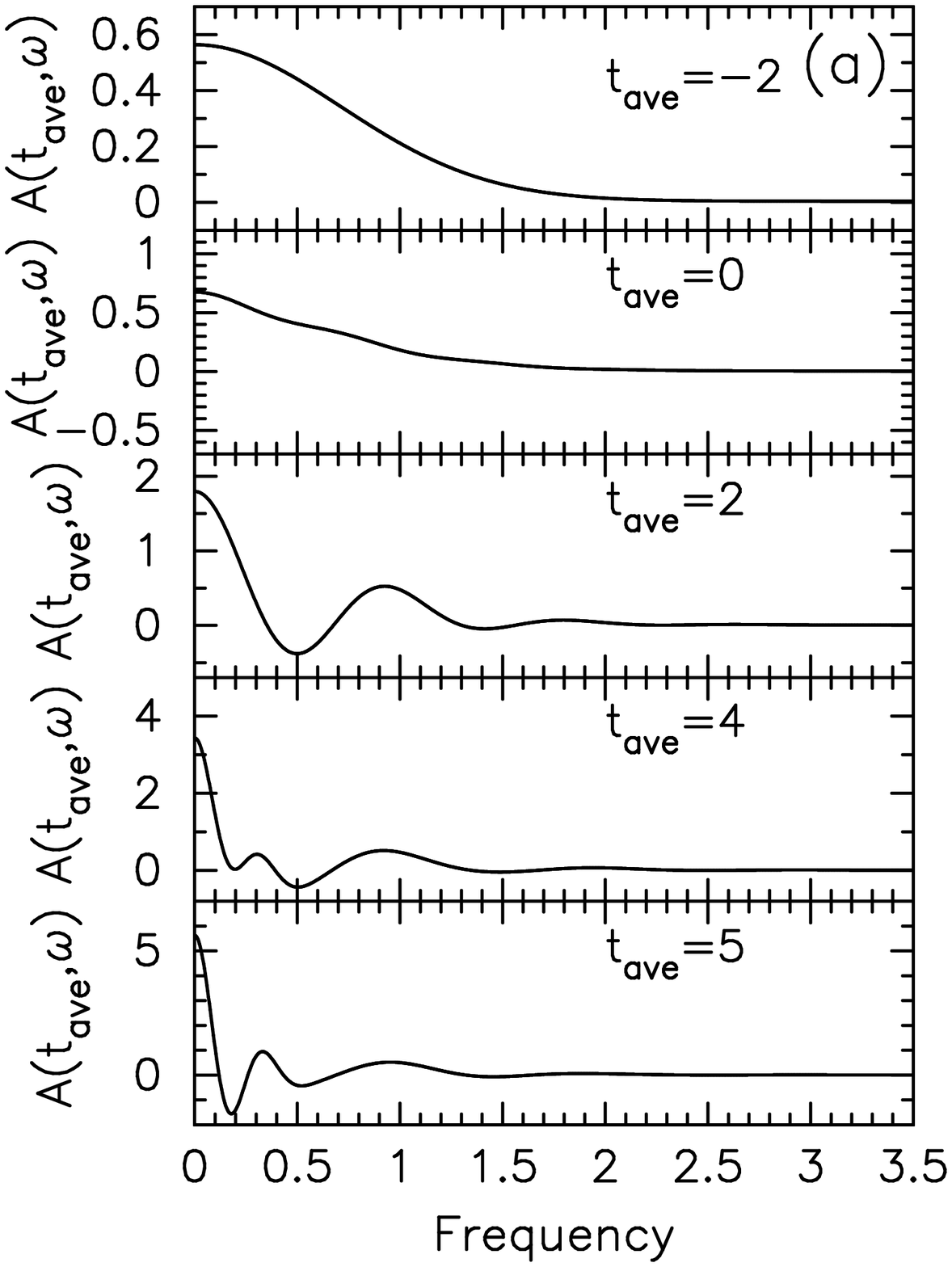}
\includegraphics[width=3.0in,angle=0]{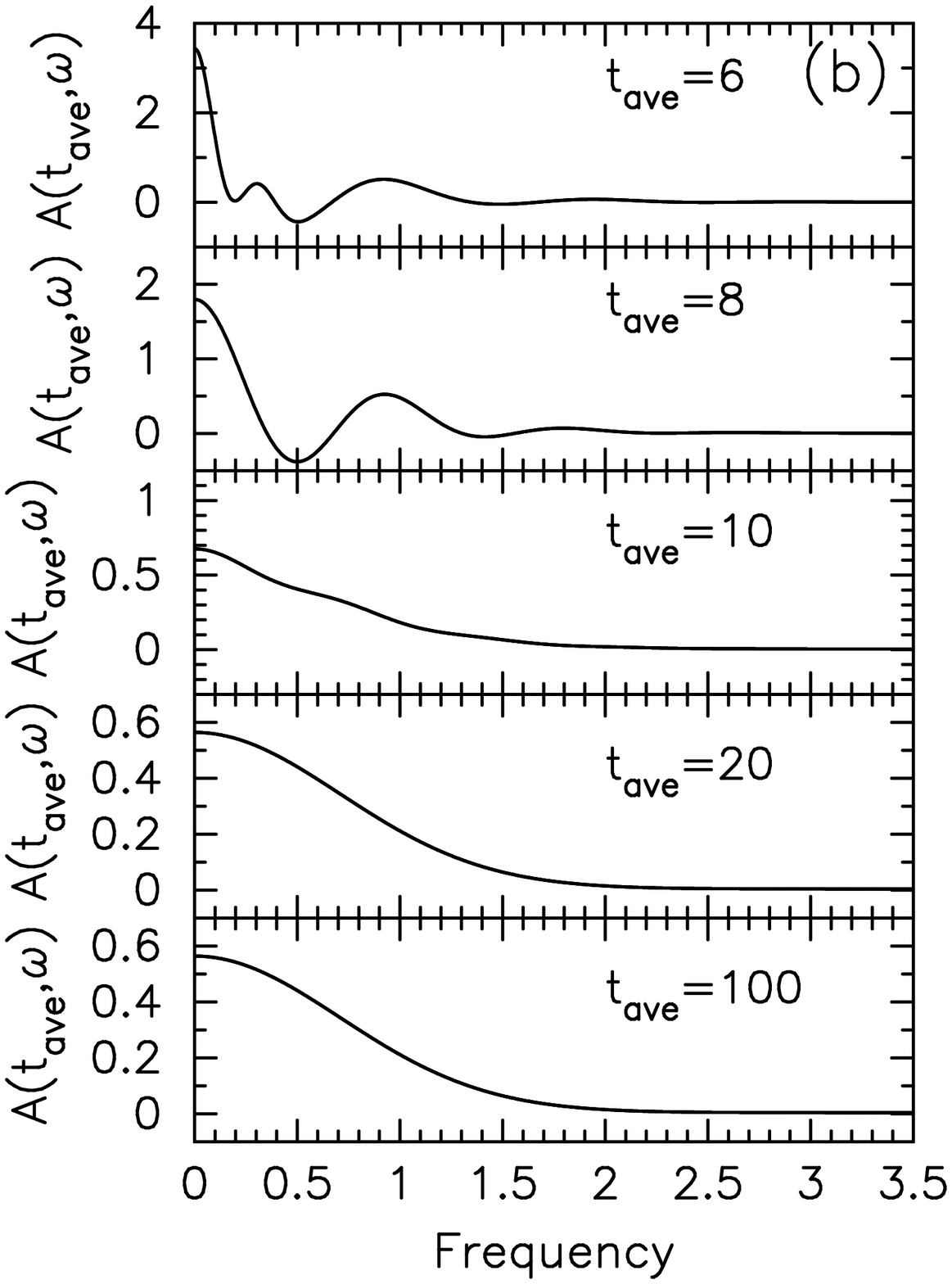}
\caption{Local DOS for the case of a sharp flat pulse with
$eaE/\hbar=1.0$, $t_{E}=10.0$, and various average times.  The horizontal scale
is the same in every panel, but the vertical scale changes in the
different panels.  By comparing
figure (a) with figure (b), one can see that the response is identical for 
times $t_{ave}$ and $t_{ave}^\prime$ that satisfy $t_{ave}+t_{ave}^\prime=t_E$.
\label{fig: sharp_pulse}}
\end{figure}

The results of the calculations are presented
in Fig.~\ref{fig: sharp_pulse} for $eaE/\hbar=1$ 
(when $eaE/\hbar$ is much smaller than 1, the
oscillations become hard to see).  The nonequilibrium DOS shows oscillating
behavior, which then decays as time increases.  The results satisfy a symmetry
relation, where the Wigner DOS is identical for $t_{ave}$ and $t_{ave}^\prime$
when $t_{ave}+t_{ave}^\prime=t_E$.

\begin{figure}[h!]
\centering
\includegraphics[width=3.0in,angle=0]{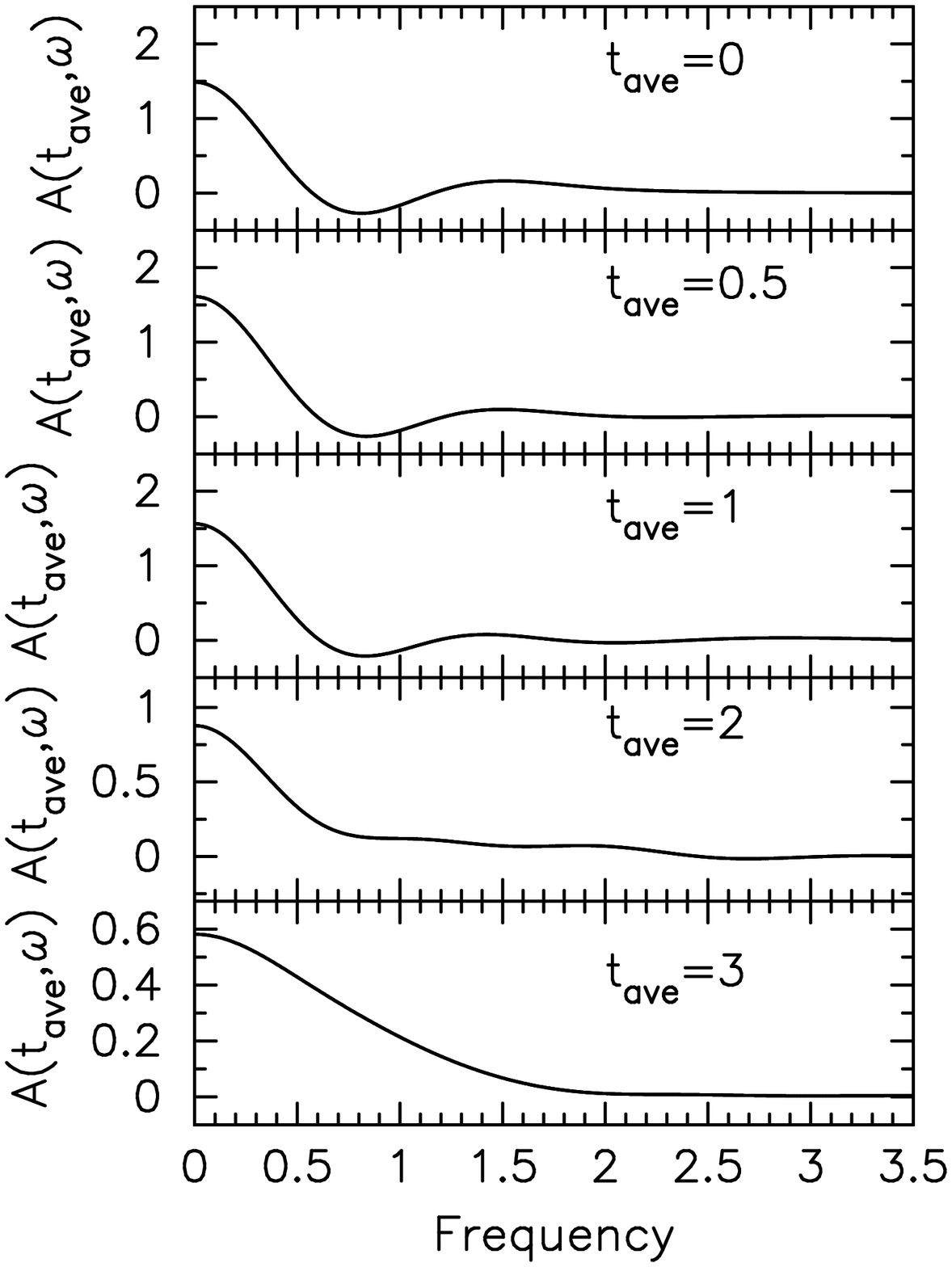}
\caption{Local DOS for the case of a smooth Gaussian pulse with
$eaE/\hbar =10.0$, $t_{E}=1.0$, and various average times.  The results are completely
symmetric between negative and positive average times, so we plot only the 
positive times here.  Note how the oscillations are already strong at 
$t_{ave}=0$, first increase slightly, then
fade away as the average time increases.
\label{fig: smooth_pulse}}
\end{figure}

We also consider the case of a smooth pulse with a smooth turn-on and  turn-off 
of the electric field: 
$A(t)=Ect_E\exp (-t^{2}/t_{E}^{2})/2$
[which corresponds to an 
electric field $E(t)=Et/t_{E}\exp (-t^{2}/t_{E}^{2})$].  This field changes
sign at $t=0$ and has it maximum amplitude at $t=\pm\sqrt{0.5}$.  The Wigner
DOS is symmetric in $t_{ave}$, so we only plot results for positive times
in Fig.~\ref{fig: smooth_pulse}.  Note that at $t_{ave}=0$ the field has been
on for a long time, so the result is far from a Gaussian.  The amplitude of
the peak in the DOS at $\omega=0$ is largest at $t_{ave}=\pm\sqrt{0.5}$, and
decays rapidly for larger times.

The proof of the symmetry relation for the Wigner DOS is rather straightforward 
to do.  If the vector potential $A(t)$ has definite parity: $A(-t)=\pm A(t)$,
then it is easy to see from Eq.~(\ref{eq: i_def}) that $I(-t_{ave},t_{rel})=
I(t_{ave},t_{rel})$ for even functions and $I(-t_{ave},t_{rel})=
I(t_{ave},t_{rel})^*$ for odd functions.  Since it is the modulus of $I$
that enters into the calculation of $A(t_{ave},\omega)$, the DOS will
satisfy the given symmetry rules.  For the case of the constant-field pulse,
we need to shift the time axis by $t_E/2$ and shift the vector potential by
$Et_E/2$ to have a vector potential that is odd in time.  The shift of the 
vector potential has no effect on the modulus of $I$, since it contributes only
a phase, while the shift in the time axis, is precisely what is needed to 
give the symmetry relation described above.  For the Gaussian pulse,
the vector potential is already an even function, and the symmetry
relation follows directly.

Note that we do not calculate the experimental probe of the reflectivity as
a function of time after the initial pulse, because this system has no
intrinsic scattering, so the optical conductivity is always a delta
function peak at zero frequency, hence we would not learn anything interesting
from such an exercise here.  It would be interesting to probe such behavior
in systems with intrinsic scattering mechanisms, to understand how the
different relaxation mechanisms can be detected.

\section{Conclusions}

We have studied the nonlinear response of Bloch electrons to an
external time varying (but spatially homogeneous) electric field by
employing an exact nonequilibrium formalism on an infinite dimensional
hypercubic lattice.
We found that the current showed Bloch oscillations, even when the electric
field was constant in time, and we derived a form for the Sharvin-like
resistance of the system.

The time-dependence of the DOS was calculated. We showed that it becomes a
Wannier-Stark ladder for long times, but the transient evolution toward those
discrete delta functions had a complex structure, that survives out to long 
times. We also examined a number of different kinds of distribution
functions, and showed that the most commonly chosen distribution
functions retained the Fermi-Dirac form regardless of the strength
of the electric field (but the local quasiparticle distribution shows complex
oscillatory behavior). For pulsed fields, we saw the transient response build
and then decay.  The amplitude of the oscillations was proportional
to the amplitude of the electric field $E$ for a wide range of field strengths,
and we needed the field to be sufficiently large 
($eaE/\hbar\sim t^{*}$) before they could be easily seen. Of course,
the oscillations decay at times larger than the pulse time.

These noninteracting Green's functions form the basis for a nonequilibrium
dynamical mean field theory, which we are currently developing to
study the nonlinear response of systems close to the Mott transition. Results
of that work will appear elsewhere.

\section*{Acknowledgments}

We would like to thank J.~Serene and V.~Zlati\'c for useful discussions.
We acknowledge support from the National
Science Foundation under grant number DMR-0210717 and the Office of Naval
Research under grant number N00014-99-1-0328.

\bibliography{fk_dmft.bib}

\end{document}